\documentclass[aps,twocolumn,showpacs,preprintnumbers,amsmath,amssymb,prb]{revtex4-1}

\usepackage{graphicx}
\usepackage{bm}
\usepackage{epsfig}
\usepackage{color}
\usepackage{hyperref}
\usepackage{subfigure}

\newcommand{\hamilton}{\hat{\mathcal{H} }}
\newcommand{\matrixX}[1]{\bm{#1}}
\newcommand{\HeThree}{\textsuperscript{3}He}
\newcommand{\HeFour}{\textsuperscript{4}He}
\newcommand{\eps}{\varepsilon}
\newcommand{\up}{\uparrow}
\newcommand{\down}{\downarrow}
\newcommand{\op}{\operatorname}

\begin{document}

\title{Ring-exchange periodic Anderson model for \HeThree{} bilayers}

\author{Jan Werner}
\email{jwerner@physik.uni-wuerzburg.de}
\author{Fakher F. Assaad}%
\affiliation{Institut f\"ur Theoretische Physik und Astrophysik, Universit\"at W\"urzburg, Am Hubland, 97074 W\"urzburg, Germany}

\date{\today}

\begin{abstract}
We present numerical results of a model calculation for the \HeThree{} bilayer system, which captures
the interplay between fast and slow dynamics of the different layers and incorporates an independent scale
for the three body ring exchange. By means of cluster dynamical mean-field theory in conjunction with continuous-time quantum Monte Carlo, we find remarkable similarities with
the experiments: a suppression of the coherence temperature upon approaching the solidification point of the first layer
accompanied by the onset of strong ferromagnetic correlations within the first layer. Based on the consistent results
for different cluster sizes we conjecture a first-order transition cutting short the Kondo breakdown, which
allows us to interpret the experimental observation of an intervening phase
preempting the quantum critical point in terms of a phase separation.
\end{abstract}

\maketitle
\section{Introduction}
Heavy fermion systems remain a subject of strong interest, even after decades of intense research activity
\cite{RMP_heavy_fermions,Nature_Heavy_Electron_Metals}. Heavy fermion superconductivity \cite{RMP_theory_unconv_SC},
quantum criticality,  the competition and
 coexistence of Fermi-liquid coherence and magnetic order
\cite{NatPhys_QC_in_HF_metals,QCP_global_phase_diagram_HF,Science_HF_and_QPT,%
ColemanFrustrationKondo,VojtaOSMott,Rev_QPT_heavy_fermions,Capponi00},
and the recent proposal of a topologically non-trivial ground state in heavy fermion compounds
\cite{Coleman_PRL_TKI,Coleman_PRB_TKI} stand as important topics.

\HeThree{} represents a unique realization of a Fermi liquid \cite{book_Quantum_liquids,Vollhardt_rev}.
In particular, layered systems of \HeThree{} have been used in different experiments
to realize unique Fermi-liquid ground states. What makes this system especially
interesting is that the relative strength of the interaction and kinetic energy
can be varied by choosing suitable substrates, and by tuning the filling.
In heat capacity measurements on \HeThree{} monolayers on graphite,
a linear dependence on temperature characteristic of a Fermi liquid
was observed \cite{heat_cap_mono_3He_graphite}. By tuning the filling,
the linear coefficient $\gamma$ can be varied, and it even diverges
at a critical filling, which marks the solidification of the monolayer.  This provides a realization of a  filling controlled Mott metal-insulator  
transition in two dimensions which  is driven by the divergence of the effective mass \cite{Imada_rev}. In  another setup with
\HeThree{} on a substrate of four layers of \HeFour{}, with two being solid and two superfluid,
on top of graphite,  a series of steps as a function of filling were observed in the $\gamma$
coefficient \cite{heat_cap_3He_on_4He}. This was attributed to the population of additional,
excited states, which results in enhanced quasiparticle interactions.

The results of experiments conducted on a bilayer system of \HeThree{} have been interpreted
in terms of heavy fermion physics \cite{3HeScience}. What makes this particular setup
so interesting is that the first layer remains fluid as the second layer is forming,
and it only solidifies at a somewhat larger filling. It is the interplay between
the slow dynamics of the strongly correlated first-layer fermions, which are close
to solidification, and the fast dynamics of the weakly correlated second-layer
fermions, which drives the heavy fermion physics in this particular system.
Hence, a heavy fermion phase is observed, where due to the Kondo effect, fermions of the different
layers form coherent quasiparticles of composite character. Accordingly, the magnetic moment of the
nearly localized first-layer fermions is screened by the delocalized second-layer fermions,
resulting in a Pauli magnetic susceptibility.
The coherence temperature $T_{\text{coh}}$ as the characteristic low energy scale
corresponds to a large mass renormalization.

These experiments clearly show that the bilayer of \HeThree{} is a unique realization
of heavy fermion physics. The specific heat below $T_{\text{coh}}$ depends linearly
on temperature, while the magnetization curve saturates to a constant.
With increasing filling the coherence scale $T_{\text{coh}}$ is suppressed due to the
density-dependent effective hybridization. By extrapolation, $T_{\text{coh}}$ is
found to vanish at a critical filling $n_C$ where the effective mass diverges;
the Kondo effect breaks down at a quantum critical point (QCP) and the first layer solidifies.
This orbital-selective Mott transition makes the first layer a local moment ferromagnet, while
the second layer is a fluid overlayer.
However, in the experiments quantum criticality is preempted by an intervening phase at filling $\langle n \rangle_I < \langle n\rangle_C$,
identified by the onset of a finite magnetization at the lowest temperature.
This indicates that the magnetic exchange coupling, which results from
three particle ring exchange processes \cite{Thouless65,Roger83,Roger98}, is becoming
the dominant energy scale, and leads to strong ferromagnetic fluctuations.

The aim of this paper is to introduce a model which captures the minimal ingredients  of the \HeThree{} bilayer:
Kondo screening, the hard core character of the \HeThree{} atoms as well as the  three particle ring exchange
which triggers ferromagnetic fluctuations in the first  layer. In Sec. \ref{sec_model_method} we will introduce the model,
and show  how to solve it within  cluster dynamical mean-field theory (DMFT).   Since we are dealing with a strong coupling problem,
we will use the continuous time hybridization expansion quantum Monte Carlo algorithm   \cite{PWernerPRL,Haule07} as a cluster solver.

In Secs. \ref{sec_results} and \ref{sec_transition}  we  detail our numerical results for various cluster sizes.
A comparison with  the experimental  data of Ref.~\onlinecite{3HeScience} as well as an interpretation of our results
is provided in  the  discussion and conclusion Sec. \ref{sec_discussion}. 
 
\section{\label{sec_model_method}Model and Method }
In a series of studies, model Hamiltonians based on the bilayer Hubbard model and periodic Anderson model have been put forward to
understand the physics of the \HeThree{} bilayers. In Refs.
\onlinecite{3He_Slave_Boson_PRL}, \onlinecite{3He_Slave_Boson_PRB} and \onlinecite{3He_Slave_Boson_Activation_Gap}
a slave boson approximation which captures the Fermi-liquid ground state but neglects local moment
formation was used. There, a vanishing of the effective hybridization was found to occur prior
to the QCP inferred from extrapolating the coherence scale.

\begin{figure}[t]
\begin{center}
\includegraphics[width=0.4\textwidth]{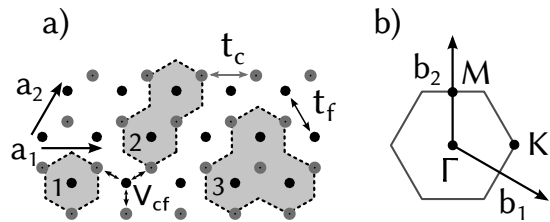}
\end{center}
\caption[]{a) Real-space arrangement of \HeThree{} atoms in the bilayer system. a$_{1,2}$ are
the lattice unit vectors. Second-layer sites (gray) are on top of and shifted with respect to
sites in the first layer (black). The double arrows indicate intralayer hopping with amplitude $t_{c,f}$
and interlayer hopping, i.e. hybridization with amplitude $V_{cf}$.
The shaded areas represent the cluster geometries employed, containing respectively
one (1), two (2) and three (3) correlated first-layer sites.
b) First Brillouin zone of the triangular lattice, with reciprocal lattice vectors b$_{1,2}$ and
high-symmetry points.}
\label{fig_lattice}
\end{figure}

A cluster DMFT study of the bilayer Hubbard model was presented in Refs.~\onlinecite{3HeModel2009,3HeModel2011}.
However, the numerical investigation was plagued by an odd-even effect, where competing ground states are realized,
which are favored by clusters with an odd and even number of sites, respectively. In the current study,
we bring forward a similar model, but with some very important differences.
We start with the same two-band Hubbard model, with the two layers arranged in
a close-packed triangular lattice as schematically shown in Fig. \ref{fig_lattice}.
In contrast to the previous works, we incorporate an independent scale for the three body ring exchange
as well as the constraint of no double occupancy. This leads to the Hamiltonian
\begin{align}
 \label{eqn_hamiltonian}
 \hamilton = & \hamilton_0 + \hamilton_I \\ 
 \hamilton_0 = & \sum_{k,\sigma} 
\begin{pmatrix} c^{\dagger}_{k,\sigma} & f^{\dagger}_{k,\sigma} \end{pmatrix}
 \begin{pmatrix}
 \varepsilon_C(k) & V(k) \\
 V(k)^{\ast} & \varepsilon_f(k)
 \end{pmatrix}
 \begin{pmatrix}
 c_{k,\sigma} \\ f_{k,\sigma}
 \end{pmatrix} \nonumber \\
 \hamilton_I = & \lim_{U \to \infty} \frac{U}{2} \sum_{i} n_{f,i}\, \bigl( n_{f,i} - 1 \bigr)
    -J \sum_{ \langle i, j \rangle } S_{f,i} S_{f,j} \notag
\end{align}
where we label first-layer fermions as $f_{k,\sigma}$ and second-layer fermions
as $c_{k,\sigma}$, with momentum $k$ and spin $\sigma$. As a result of interlayer and
intralayer hopping we obtain the dispersions $\varepsilon_c(k)$ and $\varepsilon_f(k)$
and a momentum dependent hybridization $V(k)$:
\begin{align}
\label{eqn_dispersion}
\eps_c(k) &= -2 t_c \gamma(k)\\
\eps_f(k) &= E_f-2 t_f \gamma(k) \notag\\
V(k) &= V_{cf} \sqrt{3+2 \gamma(k)} \notag\\
\gamma(k) &= \cos(k \cdot \text{a}_1) + \cos(k \cdot \text{a}_2) + \cos(k \cdot(\text{a}_1-\text{a}_2)) \notag.
\end{align}

$E_f$ is an energy offset which models the difference in binding energies of the first and second layer.
The hard-core constraint of the \HeThree{} atoms is taken into account by the limit $U \to \infty$,
which allows us to project out all states with double occupancy. The three body ring exchange
gives rise to an effective ferromagnetic coupling between nearest-neighbor spins \cite{Roger83}.
Hence, to explicitly include the ring exchange, we introduce a Heisenberg spin-spin exchange
interaction between neighboring sites $\langle i, j \rangle$, with a ferromagnetic coupling constant $J > 0$.

Note that we neglect the effect of correlations in the second layer, which is justified as long the filling of this layer
is significantly smaller than 1, i.e. half-filling, such that correlations result at most in a mild renormalization of the band dispersion.

Our model hence contains relevant magnetic interactions for \HeThree{}. In particular, since $U \to \infty$,
superexchange is blocked. On the other hand the Ruderman-Kittel-Kasuya-Yoshida (RKKY) interaction will be dynamically generated
and will compete with the ring exchange.

The concrete parameter values we choose are $t_f = t_c = t$, $V_0/t=0.5$, $J/t=0.3$ and $E_f/t = -4.0$.
We study the model in the grandcanonical ensemble as a function of the chemical potential $\mu$, which the
density couples to, and the inverse temperature $\beta = 1/T$.

To make this model accessible to numerical simulations, we use the cellular DMFT\cite{CDMFT01},
which is a real-space cluster extension to the dynamical mean-field theory \cite{RevDMFT96}, and
hence yields results in the thermodynamic limit.
In this scheme the lattice self-energy is approximated to be independent of momentum in the reduced Brillouin zone of the super cell,
while it retains its dependence on frequency. This approximate self-energy is obtained from an effective cluster of $N_c$ correlated sites
embedded in a non-interacting bath described by the hybridization function $\matrixX{\Delta}(i \omega)$.
This quantity connects the auxiliary cluster problem with the original lattice problem via the self-consistency condition
\begin{equation}
\label{eqn_Cdmft}
\matrixX{G}_{imp}(i \omega) = \frac{N_c}{N}\sum \limits_{\tilde{k}} \matrixX{G}_{latt}(\tilde{k}, i \omega)
\end{equation}
between the impurity Green's function $\matrixX{G}_{imp}$ and the local lattice Green's function
\begin{align}
\matrixX{G}_{imp}(i \omega)^{-1} & = \matrixX{i \omega_m}-\matrixX{t}_f-\matrixX{\Delta}(i \omega) - \matrixX{\Sigma}(i \omega) \\
\Bigl [ \matrixX{G}_{latt}(\tilde{k}, i \omega) \Bigr ]_{i j} & = \frac{1}{N_c} \sum \limits_K G_{latt}(\tilde{k}+K,i \omega) e^{i K \cdot (R_i-R_j)}. \notag
\end{align}
All quantities are matrices in cluster space and, in principle, depend on spin. As we will limit
ourselves to the paramagnetic, symmetry-unbroken state, we neglect the spin index.
$\tilde{k}$ is a momentum in the reduced Brillouin zone, while $R$ and $K$ are the positions
of the cluster sites in real space and reci\-procal space, respectively. This method takes dynamical
fluctuations fully into account, thus capturing the physics of Kondo screening, but it also
incorporates non-local correlations to describe the magnetic exchange. It is thus very well
suited to study models with local and short-ranged correlations \cite{RMP_Quantum_cluster}.
It is precisely the competition between these two interactions which determines the behavior
of the system, and which is at the heart of heavy fermion physics.
The relatively large coordination number ($Z=6$) of the lattice at hand means that the
approximation of cluster DMFT can yield reasonable results.

To obtain the self-energy $\matrixX{\Sigma}(i \omega)$ within each step of the self-consistency loop, we solve the effective cluster
problem by means of continuous time hybridization expansion quantum Monte Carlo \cite{PWernerPRL,Haule07}.
This numerically exact method systematically samples diagrams from the series expansion
of the partition function $Z$ corresponding to the effective action $S_{eff}$,
\begin{align}
S_{eff} &= S_c + \int \limits_{0}^{\beta} d \tau\, d \tau'\, \sum f^{\dagger}(\tau) \matrixX{\Delta}(\tau - \tau') f(\tau')\\
\frac{Z}{Z_{loc}} &= \sum \limits_{n=0}^{+\infty} \sum \limits_{\{\alpha,\alpha'\}} \frac{1}{n!} \int \limits_{0}^{\beta} d \tau_1\, d \tau'_1 \cdots d \tau_k\, d \tau'_k \\
\times & \left \langle T_{\tau} f^{\dagger}_{\alpha_1}(\tau_1) f_{\alpha'_1}(\tau'_1) \cdots f^{\dagger}_{\alpha_n}(\tau_n) f_{\alpha'_n}(\tau'_n) \right \rangle_{loc} \notag \\
\times & \frac{1}{n!} \det \left [ \Delta_{\alpha_i \alpha_j}(\tau_i-\tau'_j)\right ]. \notag
\end{align}
$S_c$ denotes the contribution of the local impurity, and we used the
shorthand notation \mbox{$f = (f_{\alpha}) = (f_{1,\up},f_{1,\down}\dots,f_{N_c,\up},f_{N_c,\down})$}.
The operator expectation value $\left < \cdots \right >_{loc}$ is taken with respect to the cluster part of the Hamiltonian.

This method is very efficient in the regime of strong interactions, and is able to treat arbitrary local
(restricted to the cluster) interactions such as Hund's rule coupling or ring exchange. The drawback is that
the numerical effort scales exponentially with the number of cluster sites or orbitals \cite{CTQMC_RMP}.
In principle, by exploiting symmetries, such as the total particle number $N$, the $z$-component of the
total spin $S_z$ \cite{Haule07}, and the projected singlet quantum number \cite{ParraghConservedQuantities},
and thereby splitting the matrices into blocks, one can reduce the computational complexity.
However, in our case, due to the hopping between cluster sites, this approach is limited to $N$ and $S_z$.

In addition, while absent for single-site calculations, in effective cluster simulations for cluster DMFT there is a sign problem,
which prohibits us from simulating clusters with more than three correlated $f$-sites at sufficiently low temperatures.

However, since the effective nearest-neighbor magnetic exchange is explicitly included in the Hamiltonian \eqref{eqn_hamiltonian},
even small clusters already include important aspects of the ring exchange physics. Therefore, even
a cluster of size $N_c=2$ represents a significant improvement over single-site DMFT, where
ring exchange is completely neglected.

In the following we present our results with plots for $N_c=2$ and $N_c=3$ side-by-side, where appropriate. However, to make the discussion of the results
more coherent, in the text we focus on the case $N_c=3$.
\section{\label{sec_results}Fermi Liquid Regime}
\begin{figure}[t]
\includegraphics[width=0.5\textwidth]{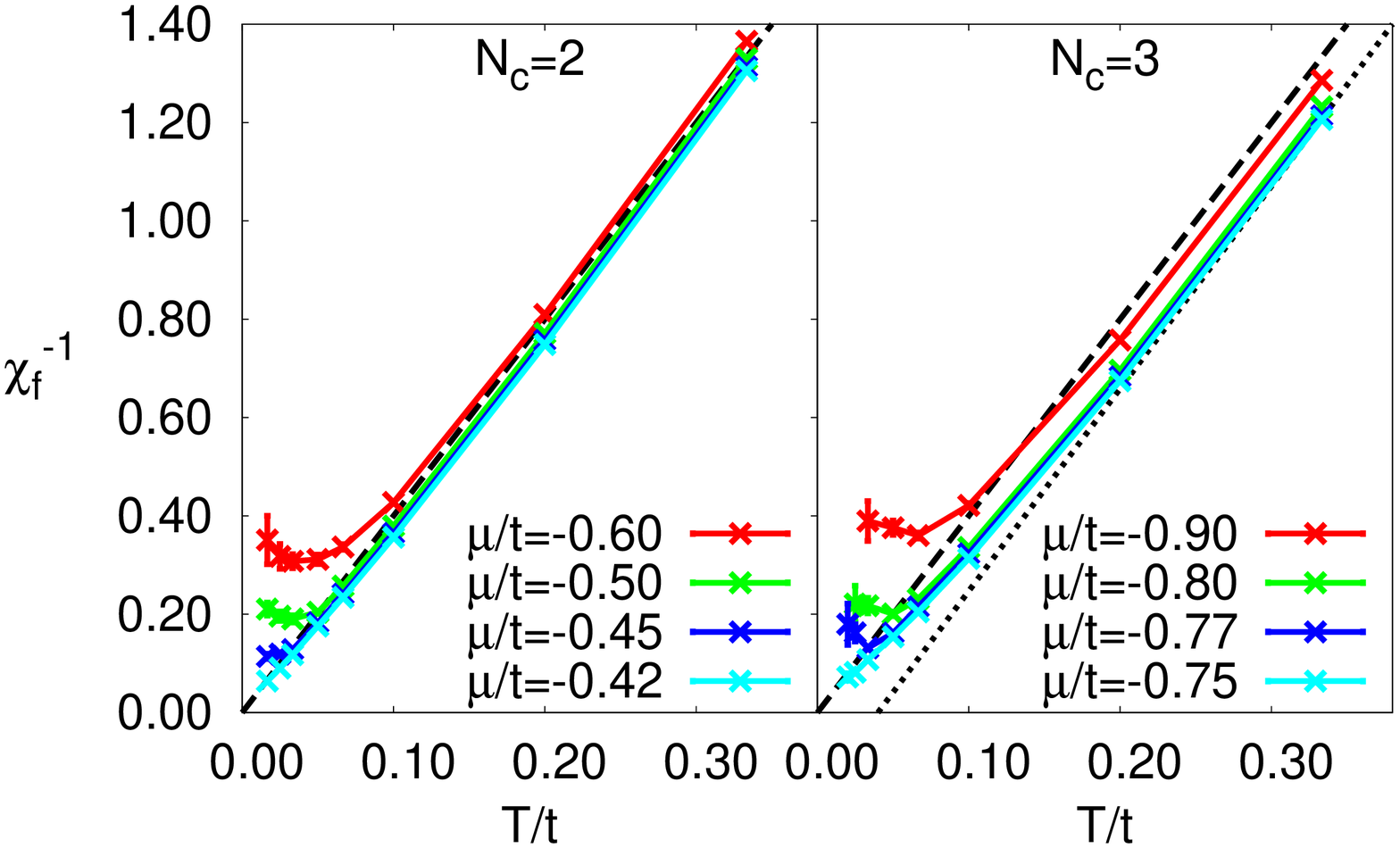}
\vspace{-5mm}
\caption{Transverse static $f$-spin susceptibility obtained from the cluster. In the local-moment regime, the layers
are uncoupled, with a behavior according to a Curie-Weiss law, while below $T_{\text{coh}}$ the $f$-spins
are screened and exhibit a Pauli susceptibility. The black dashed
line is the Curie susceptibility ($\theta=0$), while the dotted line
corresponds to $\theta=0.04t$.}
\label{fig_chi}
\end{figure}

Starting at less than half filling of the first layer, the hybridization with the delocalized
second-layer fermions leads to a crossover from an incoherent regime at high temperatures
to a regime of coherent heavy quasiparticles, which takes place at a
characteristic scale, the coherence temperature $T_{\text{coh}}$.
This crossover can be seen in the inverse of the transverse static susceptibility of the $f$-spins,
\begin{equation}
\chi_f = \chi_f(i \nu=0) = \frac{1}{V}\, \left. \frac{d M_f}{d B} \right|_{B=0}.
\end{equation}
Here we have approximated the lattice susceptibility by the cluster susceptibility,
\begin{equation}
\chi_f(i \nu) \approx \chi^{(C)}_f(i \nu) = \frac{1}{N_c} \sum \limits_{i,j=1}^{N_c} \int \limits_{0}^{\beta} d \tau\,\left \langle S_i^{z}(\tau) S_j^{z}(0) \right \rangle.
\end{equation}
The static susceptibility is shown in Fig. \ref{fig_chi}.
At high temperatures, the susceptibility follows a Curie-Weiss law $\chi_f \sim \frac{1}{T-\theta}$,
which is the expected behavior for local moments. The ferromagnetic coupling $J$ leads to
an increase in the susceptibility compared to the case of isolated moments ($\theta=0$),
and thus gives rise to a finite Weiss temperature $\theta > 0$.
For $\mu/t \leq -0.77$, the susceptibility clearly deviates from the Curie-Weiss behavior
at the lowest temperatures. Instead $\chi_f$ extrapolates to a finite value as $T \to 0$.
This corresponds to a complete screening of the $f$-moments by the Kondo effect due
to spin-flip scattering of the $c$-fermions off the $f$-moments, and it is precisely the
Pauli paramagnetism expected for a Fermi liquid.
Due to a severe negative sign problem, for $\mu/t \leq -0.80$ we are restricted
to inverse temperatures $\beta t \leq 40$. However, this is below $T_{\text{coh}}$, so that
the crossover to the Fermi-liquid state can be nicely resolved.
For $\mu/t \geq -0.75$, the situation seems to change rather abruptly. $\chi_f$ approaches the Curie-law behavior
at the lowest temperature attainable; that is the local moments are not screened.

\begin{figure}[t]
\includegraphics[width=0.5\textwidth]{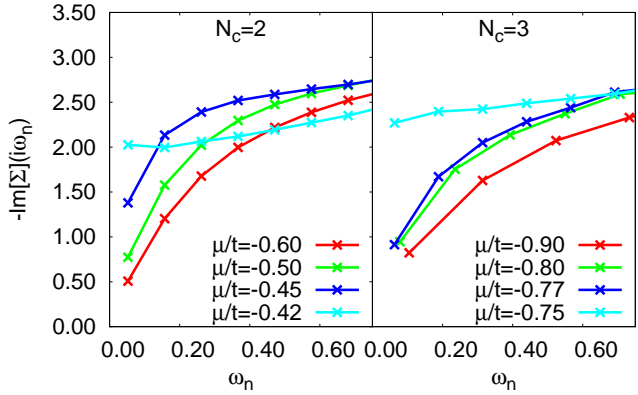}
\vspace{-5mm}
\caption{Imaginary part of the local self-energy for the clusters of size $N_c=2$ and $N_c=3$, and
at the lowest temperatures attainable. For $\mu/t \leq -0.45$ ($\mu/t \leq -0.77$), it is linear
at small frequencies, which signifies the existence of a Fermi liquid at low temperatures.}
\label{fig_sigma}
\end{figure}

On the real frequency axis and for the generic Fermi-liquid state, the low frequency behavior of the self-energy  is characterized by 
$\op{Im}[\Sigma( \omega)] \sim  \omega^2$, and  $\op{Re}[\Sigma( \omega)] \sim  a +  b \omega$.     For Matsubara   frequencies, this implies that 
$\op{Im}[\Sigma(i \omega_m)] \sim  \omega_m$. 
In Fig. \ref{fig_sigma} the low-frequency imaginary part of the local self-energy
is plotted.
For smaller values of $\mu/t$ there is a clear signature of a Fermi-liquid ground state.
Beyond a certain threshold value, an extrapolation to zero with $\omega \to 0$ is clearly not possible.

\begin{figure}[t]
\includegraphics[width=0.5\textwidth]{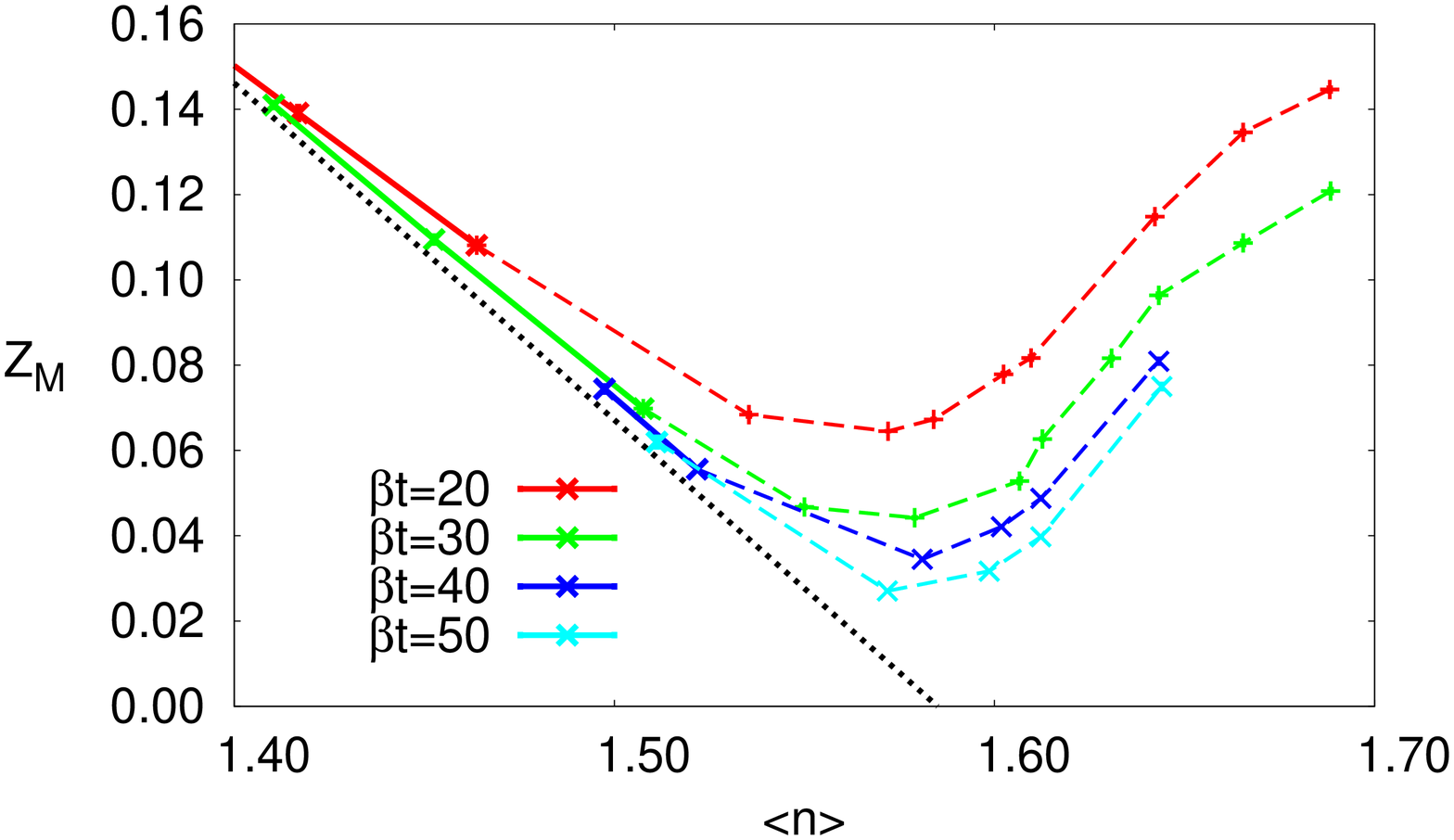}
\vspace{-5mm}
\caption{Quasi-particle residue $Z$ approximated by $Z_M$ as defined in \eqref{eq_ZM} for different temperatures,
derived from the results for $N_c=3$.
The dashed lines imply that either the imaginary
part of the self-energy is not yet linear in frequency because of the
high temperature or $Z$ is not well defined because the self-energy is
not linear at all above the critical filling. The dotted line is a guide
to the eye to show the apparent vanishing of $Z$ at a QCP.}
\label{fig_Z}
\end{figure}

The linear coefficient of the self-energy at small frequencies is related to the coherence temperature
$T_{\text{coh}}$, quasiparticle residue $Z$, and effective mass ratio $m^*/m$ via
\begin{equation}
\label{eq_Z}
T_{\text{coh}}^{-1} \sim Z^{-1} = \frac{m^*}{m} = 1-\left . \frac{d\, \Sigma(\omega)}{d\, \omega} \right |_{\omega=0},
\end{equation}
which we approximate using the first Matsubara frequency value via
\begin{equation}
\label{eq_ZM}
Z^{-1} \approx Z_M^{-1} = 1-\frac{\text{Im} [\Sigma(i \pi T)]}{\pi T}.
\end{equation}
Within the Fermi-liquid regime, the onset of coherence shifts to lower
temperatures with increasing chemical potential. This gradual suppression
of the coherence scale and quasiparticle residue, or equivalently the
increase of the quasiparticle effective mass, results from the renormalization
of the effective hybridization due to the increasing filling.
The evolution of $Z_M$, i.e. the inverse effective mass,
is shown in Fig. \ref{fig_Z} versus the total filling $\langle n \rangle  = \langle  n_f +n_c \rangle $.
At $n \leq 1.5$ the results for $Z_M$ converge at low temperatures, $Z_M(T \to 0) \to Z$,
once the coherent Fermi-liquid state is established. As a function of filling,
the quasiparticle residue changes approximately linearly. With increasing $\mu$, $Z$ is increasingly suppressed
and by extrapolation seems to vanish at a critical filling $\langle n \rangle_C \approx 1.58$ at a QCP.

\section{\label{sec_transition}  Magnetic crossover }

\begin{figure}[t]
\includegraphics[width=0.5\textwidth]{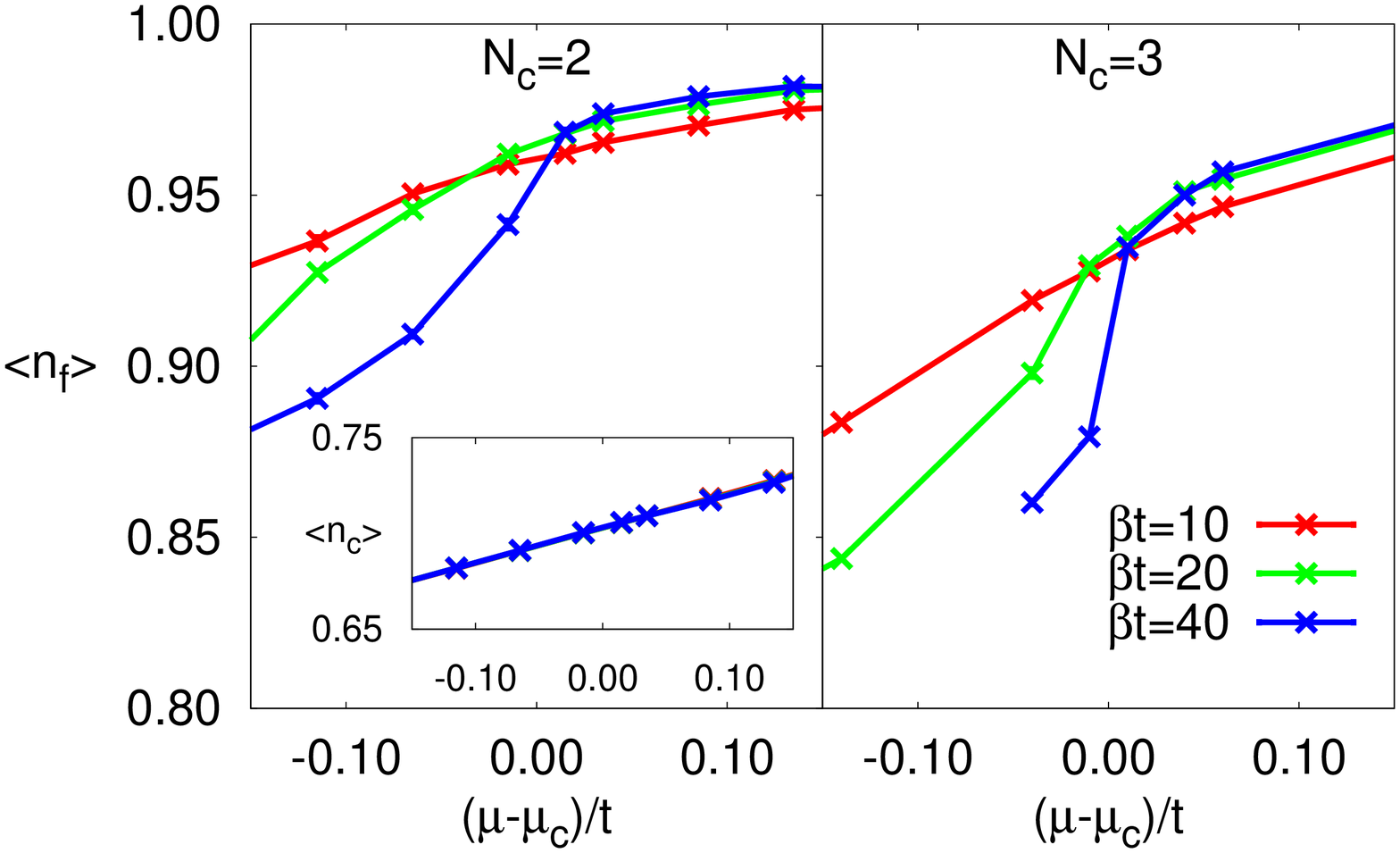}
\vspace{-5mm}
\caption{$f$-orbital occupation number for both cluster geometries. At lower temperature, the occupation exhibits
a strong increase upon crossing $\mu_C$, which is even steeper for the larger cluster. The $f$-filling
exhibits a strong temperature dependence for $\mu < \mu_C$, but is only weakly temperature dependent beyond $\mu_C$.
The inset shows the filling of the second layer for $N_c=2$, which is much smaller than 1.
Thus, neglecting correlations for the $c$-fermions is well-justified. In contrast to $\langle n_f\rangle$, the $c$-filling
has almost no temperature dependence.}
\label{fig_nf}
\end{figure}

While increasing $\mu$, the behavior of the observables considered above changes abruptly at a certain threshold value of $\mu$,
with clear deviations from the Fermi-liquid results. To learn more about what happens there, we
show in Fig.~\ref{fig_nf} the evolution of the $f$-orbital occupation as a function of chemical potential
$\mu/t$ and inverse temperature $\beta t$.
With decreasing temperature a rather strong increase of the $f$-orbital occupation develops in a very narrow window of $\mu$-values.
We define a characteristic value of the chemical potential $\mu_C$ by the largest step between subsequent values of $\mu$.

\begin{figure}[t]
\includegraphics[width=0.5\textwidth]{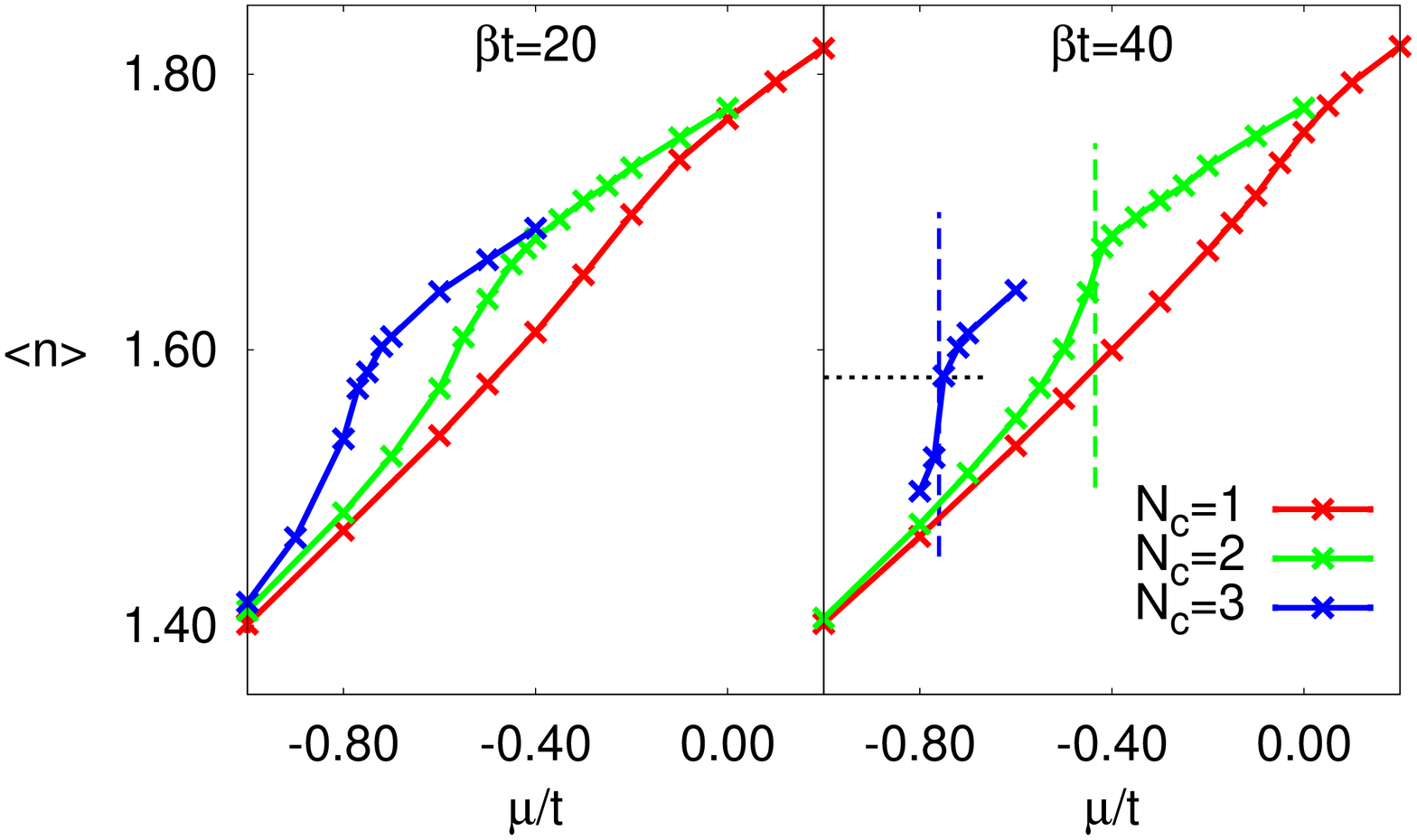}
\vspace{-5mm}
\caption{Total occupation number for $N_c=1-3$, at two inverse temperatures $\beta t$. The dashed lines indicate the
location of $\mu_C$ for $N_c=2\text{ and }3$, while the black dotted line is the critical filling $\langle n\rangle_C \approx 1.58$ obtained
for $N_c=3$.}
\label{fig_n}
\end{figure}

In addition, we compare in Fig. \ref{fig_n} the total particle number for all cluster sizes at two different temperatures.
Looking at the trend from $N_c=1-3$, the increase around $\mu_C$ becomes steeper for larger clusters and lower temperatures, but stays continuous.
It is interesting to note that the putative QCP at filling $\langle n \rangle_C$ seems to be located more or less inside the step in occupation.

\begin{figure}[t]
\includegraphics[width=0.5\textwidth]{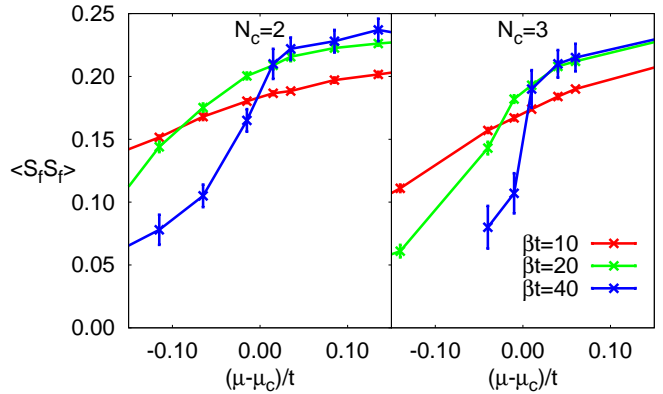}
\vspace{-5mm}
\caption{Spin-spin correlation between neighboring $f$-sites, obtained from the cluster. While there is little correlation in the incoherent
and heavy fermion phase, beyond $\mu_C$ the spins are almost fully polarized.}
\label{fig_sisj}
\end{figure}

Next, we look at the nearest-neighbor spin-spin correlation between $f$-orbitals,
\begin{equation}
\bigl \langle S_f S_f \bigr \rangle = \frac{1}{N_p} \sum \limits_{\langle i,j \rangle} \bigl \langle S_{f,i} S_{f,j} \bigr \rangle,
\end{equation}
normalized by the number of distinct pairs $\langle i,j \rangle$ of neighboring sites $N_p$, which is shown in Fig. \ref{fig_sisj}. For smaller filling,
the observed correlation extrapolates to a small value as $T \to 0$. This is consistent
with the interpretation of a Fermi-liquid state, where the moments are predominantly screened
by the second-layer light fermions.
Beyond the point $\mu=\mu_C$, we observe a completely different behavior:
a large correlation which seems to grow and eventually saturate at lower temperatures. Apparently,
for $\mu > \mu_C$ there are strong ferromagnetic fluctuations; that is, a parallel
alignment of neighboring $f$-spins is favored.

\begin{figure*}[t]
\includegraphics[width=\textwidth]{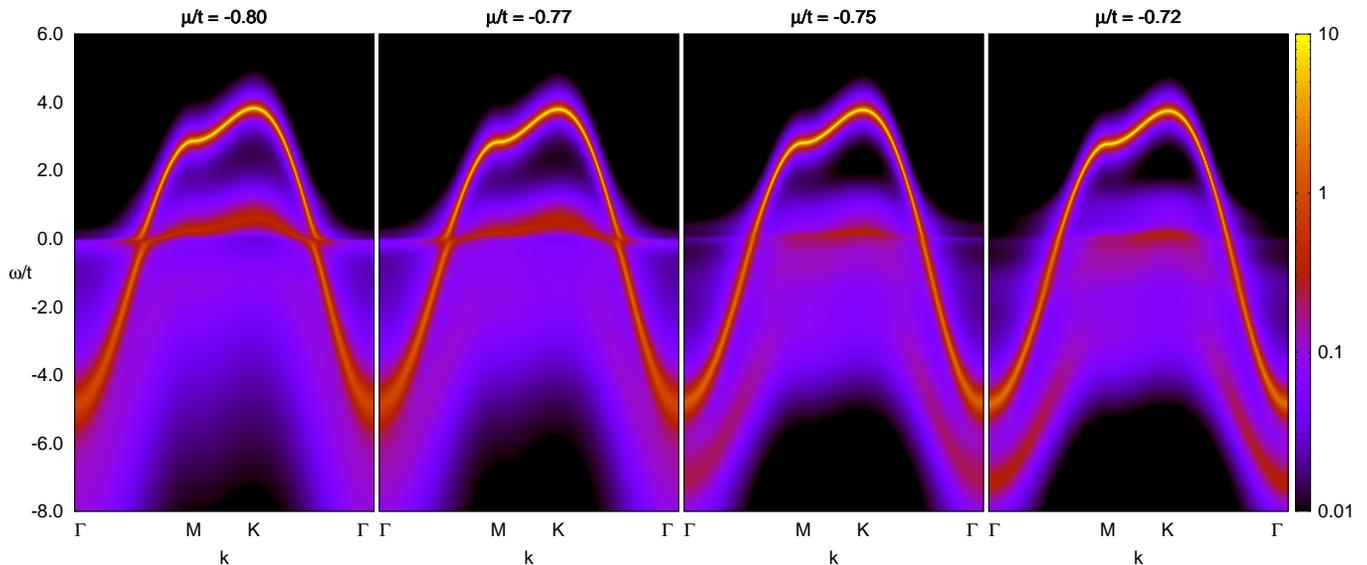}
\vspace{-5mm}
\caption{Spectral function $A(k,\omega)$ for $N_c=3$, at inverse temperature $\beta t = 40$.
For $\mu/t \leq -0.77$ the heavy-fermion bands are clearly visible,
while for larger $\mu$ there is little spectral weight left
at the Fermi energy.}
\label{fig_Akom}
\end{figure*}

Finally, we show in Fig. \ref{fig_Akom} the momentum-resolved spectral function obtained
by analytical continuation of the Green's function from Matsubara frequencies
to real frequencies \cite{KBeachMaxent}, which we obtain from the periodized
$f$-orbital self-energy $\Sigma(k,i \omega)$
\begin{align}
A(k,\omega) &= -\pi^{-1} \op{Im} \bigl [ \op{Tr} \matrixX{G}(k,\omega) \bigr ]\\
\matrixX{G}(k,i \omega) &= 
\begin{pmatrix} i \omega + \mu-\eps_c(k) & -V(k) \\
-V(k)^{\ast} &  i \omega + \mu -\eps_f(k)-\Sigma(k,i \omega)
\end{pmatrix}^{-1} \notag
\end{align}

In the two left panels in Fig. 8, where $\mu<\mu_C$, the hybridization between the two bands is clearly visible,
which leads to the heavy-fermion band close to the Fermi level, and a small gap at the points of avoided crossing.
With the suppression of the coherence scale, the heavy band becomes even flatter.
At $\mu>\mu_C$, the hybridization is almost completely gone. The $c$-band
goes straight through the points, where previously a crossing was avoided by the hybridization.
The spectral weight from the heavy-fermion band is largely gone, because it is transferred to the broad,
incoherent features far below the Fermi level.

\section{\label{sec_discussion}Discussion  and Conclusions}
Our model captures nicely the Fermi-liquid regime, where we recover the crossover from the incoherent regime
to the Fermi-liquid state of strongly renormalized quasiparticles. The coherence temperature $T_{\text{coh}}$, which marks
the onset of Fermi-liquid behavior, is suppressed by increasing the filling $\langle n \rangle $.    Extrapolating $T_{\text{coh}}$
for fillings $\langle n \rangle  < 1.5 $, where results are converged in temperature,  defines a  quantum critical
point at $\langle n \rangle_C  = 1.58$  where $T_{\text{coh}}$ vanishes and the Kondo effect breaks down.
	
	As the coherence temperature drops, or, equivalently, the  effective mass grows, the  model system becomes increasingly susceptible  to  ferromagnetic fluctuations triggered by  the three site ring exchange term  which competes with Kondo screening. 
Indeed, our numerical results show a crossover to a state with 
robust ferromagnetic correlations  within the first  layer. This is accompanied by a strong increase in the occupation of the first layer. 
At the same time, the high temperature spin susceptibility shows a Curie-Weiss law with  a robust  positive Weiss constant, which is appropriate  for a ferromagnetic transition. 
At low temperatures,  and on our finite size clusters,  the Curie-Weiss law gives way to a finite a Pauli spin susceptibility.
This low temperature behavior is  clearly very sensitive to the cluster size. Indeed, on any finite lattice
the ferromagnetic  symmetry broken state will not appear and at low enough temperatures one expects
Kondo screening of the local moments of the first layer by the second layer light fermions. 

It is important to note that the crossover observed in the occupation as a function of cluster size becomes sharper and sharper and  seems to 
extrapolate  to  fillings lower than  $\langle n \rangle_C  = 1.58$, as obtained from  the coherence temperature.   The  above allows for an interpretation in the 
infinite cluster size limit which is in remarkable agreement with the experimental data of Ref.~\onlinecite{3HeScience};   the QCP is preempted by a first-order  transition to a ferromagnetic  state. 
The transition happens when the polarized state becomes more favorable than the Fermi-liquid state stemming from Kondo screening of the local moments of the first  layer. 
This is the case when $T_{\text{coh}}$, i.e. the energy of formation of quasiparticles, becomes smaller than
the energy the system gains from a parallel alignment of the $f$-spins, which is the effective coupling $J^{\ast}$.
$J^{\ast}$ is composed of two contributions, $J^{\ast} = J + J_{\text{eff}}$. The first part is the
bare coupling $J$, while the second is the effective  exchange $J_{\text{eff}}$
dynamically  generated by, e.g.,  the RKKY interaction.  With increasing filling,
$T_{\text{coh}}$ is suppressed so that   the ferromagnetic energy   overcomes  the coherence
scale, and the polarized state becomes the new ground state.


\begin{figure}[t]
\includegraphics[width=0.4\textwidth]{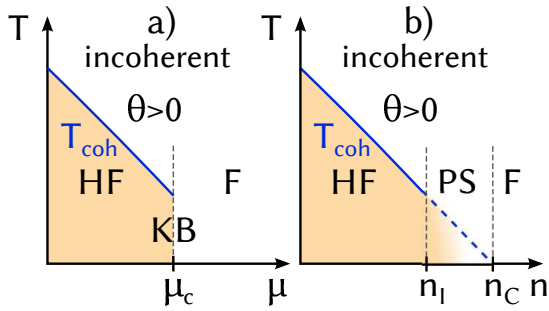}
\caption{Schematic phase diagram of the \HeThree{} bilayer as a function of a) chemical potential,
and b) total filling. The incoherent phase with a positive (ferromagnetic) Weiss temperature $\theta$
crosses over to the coherent heavy fermion state (HF) around the coherence temperature $T_{\text{coh}}$.
At the critical chemical potential $\mu_C$, the Kondo effect breaks down (KB) due to the 
first-order transition to the ferromagnetic state (F). As a function of filling,
for $\langle  n\rangle>\langle  n\rangle_I$ a phase separation (PS) is observed, while extrapolation to a vanishing coherence scale
yields the putative QCP at $\langle  n\rangle_C$.}
\label{fig_phase_diagram}
\end{figure}
Let us now compare our  model calculations to  the experimental results of Ref.~\onlinecite{3HeScience}. On the one hand we recover the heavy-fermion state, and the decrease in the coherence
scale with increasing filling, with an apparent QCP at a critical filling $\langle  n\rangle_C$, which is obtained by extrapolation.
On the other hand, we can offer an interpretation for the appearance of an intervening phase with a finite magnetization
at a lower filling $ \langle n\rangle_I <  \langle n\rangle_C$. Since our results support the picture of  a first-order transition between competing ground states  
as a function of  $\mu$, the experimental results can be understood as a phase separation, as shown schematically
in Fig.~\ref{fig_phase_diagram}. This is because in the experiment the filling of \HeThree{} atoms is controlled  and not the chemical potential. 
Thus, the discontinuous behavior with increasing $\mu$ in the simulations is mirrored in experiment
by the appearance of ferromagnetic islands that are growing in size inside the Fermi liquid
of heavy quasiparticles.

The first-order transition brings about an abrupt breakdown of the Kondo effect.  In the ferromagnetic phase we expect the hybridization matrix element at low 
energies to  vanishes.   This essentially corresponds to an orbital-selective  Mott transition: on orbitals with dominant $f$-character (i.e. predominantly first layer)  we observe 
a magnet of ferromagnetically coupled, local moments, while  on orbitals with dominant $c$-character (i.e. predominantly second layer)  we witness a  light Fermi liquid. 

\begin{acknowledgments}
We would like to thank J. Saunders for discussions. Funding from the DFG under the Grant No. AS120/6-2 (Forschergruppe FOR 1162) and from the Elite Network of Bavaria is acknowledged. 
We thank the  J\"ulich Supercomputing Centre for generous allocation of CPU time. The authors gratefully acknowledge the Gauss Centre for Supercomputing e.V. (http://www.gauss-centre.eu)
for funding this project by providing computing time on the GCS Supercomputer SuperMUC at the Leibniz Supercomputing Centre (LRZ, http://www.lrz.de).
\end{acknowledgments}

\bibliography{journal_short,3he,fassaad}
\end{document}